\documentstyle[12pt]{article}
\begin{document}
\large
\centerline{\bf D=3: Singularities in gravitational scattering of
scalar waves.}
\vskip 2 cm
\normalsize
\centerline{C. Klim\v c\'\i k}

\small \it
\centerline{Nuclear Centre, Charles University, Prague}
\centerline{Czechoslovakia}

\vskip 1 pt
\centerline{and}

\vskip 1 pt
\normalsize \rm
\centerline{P. Koln\'\i k}

\small \it
\centerline{Department of Theoretical Physics, Charles University, Prague}
\centerline{Czechoslovakia}
\vskip 1 cm

\rm \centerline{ICP classification number: 0430.}

\vskip 5 cm
\footnotesize
\noindent {\bf Abstract. }
Family of exact spacetimes of D=3 Einstein gravity interacting with
massless scalar field is obtained by suitable dimensional reduction
of a class of D=4 plane-symmetric Einstein vacua. These D=3
spacetimes describe collisions of line-fronted asymptotically null
excitations and are generically singular to the future.
The solution for the scalar field can be decomposed into the Fourier-Bessel
modes around the background solitons.
The criteria of regularity of incoming waves are found.
It is shown that the appearance of the scalar curvature singularities
need not stem from singularities of incoming waves.
Moreover, in distinction to D=4 case, for \it all \rm solutions
with regular incoming waves the final singularities are inevitable.
\newpage
\large
\S 1. Introduction

\vskip 1 cm
\normalsize
Motivations for studying the Einstein gravity coupled to source field
in D=3  are natural and were discussed many times in literature
since the interest in the subject had appeared [1, 2].
Clearly, the absence of dynamical degrees of freedom for the pure
Einstein gravity, as the crucial property of spacetimes with dimension 3,
makes possible to identify all degrees of freedom of interacting systems
with those of the matter field [1, 2].
This feature is obviously attractive from the point of view
of quantum theory since one may identify the matter field degrees
of freedom and attempt to quantize them.
Such D=3 quantum gravity should elucidate a lot of intriguing
questions arising in attempts to quantize gravity in D=4.
One expects, in particular, that quantum corrections to the classical
dynamics will be substantial near the ubiquitous singularities of
classical general relativity.
What, then, is the status of singularities in the quantum theory?

Consulting the literature on D=3 gravity it may seem that the curvature
singularities are somewhat less ubiquitous and, in general, milder
than their counterparts in D=4. Indeed, much effort was devoted to the study
of spacetimes describing gravitating (moving) point masses
[2, 3], in which only mild quasiregular singularities appear, located
at the world lines of the particles. Such spacetimes are constructed
by appropriate glueing of flat pieces in order to satisfy the Einstein
equations with the point-like sources and they are distinguished
rather by their global topological properties than by their local
structure. The interaction of massive particles was actually studied by
various methods in series of papers even at the quantum level [3--8],
however, the problem of singularities was not the central one in these
considerations.

We feel, however, that the local
structure of matter solutions may itself represent an interesting
problem to pursue.
Indeed, the stronger singularities (i.e. not quasiregular) are not
alien to D=3 general relativity, e.g. as the ``collapsing dust"
solution of [1] shows. Such curvature
singularities, however, then automatically
imply the necessary singularities of the matter stress tensor
in distinction to D=4 case.
This is not, to our point, a substantial flaw of D=3 general relativity,
since the singularities of a matter distribution may well develop
dynamically from a smooth nonsingular distribution of the matter
in the past. Moreover,
degrees of freedom responsible for the creation of the singularity
are exclusively those of the matter field, hence their quantization
should reveal much about a quantum status of singularities.
For a quantitative study of such project we need a tractable
example of classical formation of the curvature singularity. The
singularities occuring in classical scattering processes seem to be
the best starting point for considering their quantum picture, since,
the scattering processes are those best adressed in quantum field
theory.

As is well-known, there exist a lot of solutions in D=4 general relativity
describing collisions of gravitational plane-waves in which, generically,
the scalar curvature singularities are created in the future
[9--15]. These singularities can be physically interpreted
as arising due to mutual focusing of colliding waves. This interpretation
stems from the behaviour of a congruence of null geodesics hitting
transversally the wave-front of a \it single \rm D=4 gravitational
plane-wave. The congruence gets focused to the line after crossing
the wave [16]. The location of this focal line corresponds to the location
of curvature singularities of the full spacetime describing the
collision of the waves. In D=3 we observe a similar behaviour of
geodesics hitting the wave-front of a single line-wave (see \S 2),
hence a creation of curvature singularities in collisions of D=3
scalar field line-wave excitations may be expected.
We shall see, in what follows, that it indeed happens.
Moreover, the field equations get effectively linearized and one can
obtain a mode decomposition of the matter  scalar field around some
(solitonic) background configuration.
This feature seems to be particularly attractive from the point of view
of further quantization.

{}From the technical point of view, we obtain D=3 solutions by suitable
dimensional reduction of D=4 vacua describing collisions of collinear
plane-waves [15]. The field equation for one component of D=4
metric turns out to be identical to that for D=3 massless scalar
field\footnote{Some matter solution in D=4 were also constructed
by reduction of D=5 vacua [13].}.
Similarly, suitable combination of remaining D=4 metric components
fulfil the same equations as the components of D=3 metric. In this way,
we may readily translate D=4 vacua into D=3 matter solutions.
What differs substantially, however, are conditions for regularity
of incoming waves and a criterion for avoiding the final singularity.
We shall show that D=3 ``incoming" regularity conditions
are more restrictive. This restriction is so strong that it excludes,
in distinction to D=4 case, solutions avoiding the final singularity.

The paper is organized as follows. In \S 2 the properties of
\it single \rm line-wave solutions in D=3 are discussed.
The D=3 field equations for the colliding waves are obtained and
compared to D=4 equations in \S 3 and the class of D=4 vacuum
solutions is dimensionally reduced to give D=3 matter solutions
in \S 4. In \S 5 the criteria for avoiding an incoming parallelly propagated
non scalar curvature singularities are found and solutions fulfilling
these criteria are identified. The scalar curvature singularities
are studied in \S 6 and in \S 7 we furnish conclusions and outlook.

\large
\vskip 1 cm
\S 2. Line-waves in D=3.

\normalsize
\vskip 1 cm
Plane-wave spacetimes in general relativity in arbitrary dimension D
are metrics of the form
$${\rm d}s^2=-{\rm d}U{\rm d}V+h_{ij}(U)X^iX^j{\rm d}U^2+{\rm d}X^i
{\rm d}X_i\eqno(2.1)$$
where $U,V$ and $X^i\ (i$=1,...,D-2) are so called Brinkmann
coordinates\footnote{The metric (2.1) describes propagation
of disturbance of geometry in $Z(\equiv U-V)$ direction with velocity of
light.}
{}.
The exact form of the amplitude $h_{ij}(U)$ is given by Einstein
equations, e.g. for the vacuum case one obtains the condition
$$\sum_{i=1}^{{\rm D}-2} h_{ii}(U)=0.\eqno(2.2)$$
It is obvious  that for D=3 (where we  use the term ``line-waves"
instead of ``plane-waves") (2.2) gives $h(U)$=0, i.e. the flat metric,
as it should.

We shall not consider, for the moment, the dynamically allowed form of
$h(U)$ in D=3 but, rather, we turn to the behaviour of geodesics in the metric
(2.1). The most effective way to do it is to change
coordinates putting the metric in the so-called Rosen form
$${\rm d}s^2=-{\rm d}u\ {\rm d}v+{F^2(u)\over 2}{\rm d}x^2\eqno(2.3)$$
where, from now on, we shall consider the case D=3.
The transformation from Brinkmann to Rosen coordinates is given by
$$U=u,\   V=v+{x^2\over 2}F(u)F'(u),\   X={xF(u)\over {\sqrt2}};\eqno(2.4)$$
thus
$$h(U)={F''(u)\over {F(u)}}.\eqno(2.5)$$

The symmetries of the line-wave metric (2.1) are clearly more explicit
in the Rosen form (2.3), where the $x$-dependence drops out. There is
a drawback of Rosen coordinates, however, that they do not cover all
manifold by a simple chart as the Brinkmann coordinates do.
This fact can be shown to follow easily from
Einstein equations for scalar field
coupled to D=3 gravity. One supposes that the metric (2.3) is flat for all
$u<u_w\ (w\equiv$ wavefront ) and shows that for some $u_0>u_w$ the
transformation (2.4) breaks down as $F(u_0)$=0.

It is easy to find a congruence of null geodesics crossing the wave
front transversely. It contains lines
$$v=v_c,\   x=x_c.\eqno(2.6)$$
In Brinkmann coordinates, however, the congruence looks as follows
$$V(u)=v_c+{x^2_c\over 2}F(u)F'(u),\   X(u)={x_cF(u)\over {\sqrt2}}\eqno(2.7)$$
where $u$ may be taken as the affine parameter. Clearly, at $u_0$=0,
the family gets focused to the point $V$=$v_c$, $X$=0.

Assume that we consider a scattering of two line-waves
described by $F(u)$ and $G(v)$ respectively and propagating in the
opposite direction with sharp wave-fronts, i.e. $F(u)$=$G(v)$=$\sqrt 2$
for $u<u_w,\ v<v_w$; $G(v)$ plays role of $F(u)$ in (2.3) for the other
wave. The wave-front $W_v$ of $v$-wave will be modelled in $u$-wave
Rosen coordinates (2.3) by null geodesics with $v_c$=$v_w$ and $x_c$
varying over the $x$-axis. We draw the wave-front $W_v$ in both Rosen
(Fig.1a) and Brinkmann (Fig.1b) coordinates of $u$-wave.
One may draw Fig.1a with $x$ coordinate suppressed as in Fig.2. There
we can see that the crossing wave-front $v_w$=0 cuts out the part of
Rosen spacetime indicated by hatch.
As Fig.1b shows, we cannot continue the remaining part over $u>u_0$ without
crossing wave-front $v_w$=0\footnote{This continuation for $single$
line-wave is possible showing that the $u_0$ is only a coordinate
singularity. In the case of colliding waves, however, $u_0$ are truly singular
points of the spacetime.}.
The apparent half plane $u=u_0$, $v<0$
in Fig.2 is in fact the half line in the Brinkmann coordinates,
playing the role of a seam in the crossing wave-front. Therefore, we expect
on the basis of presented considerations, that the spacetime
describing the colliding waves will look in the Rosen form as in Fig.3a
where $u$=$u_0$ and $v$=$v_0$ are seams on the wave-fronts of colliding waves
and represent themselves singularities of the spacetime. The character
of these singularities depends on the amplitudes $h(U)$ and $\widetilde h(V)$
of the incoming waves, as we shall see later.

We feel, however, that the waves with sharp wave-fronts are
not very physically
realistic. Therefore, we expect the colliding solutions as drawn in Fig.3b,
which may be considered as the asymptotic version of Fig.3a. The jagged lines
in the null infinities are singular points of the spacetime.

\vskip 1 cm
\large
\S 3. Colliding waves spacetimes.

\vskip 1 cm
\normalsize
As the analysis of \S 2 suggests, we shall seek for a metric for colliding
waves in the Rosen coordinates starting with a natural ansatz
$${\rm d}s^2=-{\rm e}^{-K(u,v)}{\rm d}u\ {\rm d}v
             +{\rm e}^{-N(u,v)}{\rm d}x^2.\eqno(3.1)$$
This ansatz comes from analogy with higher dimensional case and we
restrict ourselves to the metrics with one spacelike Killing vector.

The field equations with  scalar field as a source then read
$$N_{uu}-{1\over 2}N^2_u+N_uK_u=2\kappa\phi_u\phi_u\eqno(3.2a)$$
$$N_{vv}-{1\over 2}N^2_v+N_vK_v=2\kappa\phi_v\phi_v\eqno(3.2b)$$
$$K_{uv}=\kappa\phi_u\phi_v\eqno(3.2c)$$
$$N_{uv}={1\over 2}N_uN_v\eqno(3.3)$$
$$4\phi_{uv}-N_u\phi_v-N_v\phi_u=0\eqno(3.4)$$
where $$\kappa=8\pi G.\eqno(3.5)$$

The equation (3.3) can easily be integrated to give
$$N=-2\ln\bigl(1-f(u)-g\,(v)\bigr)\eqno(3.6)$$
where $f(u)$ and $g\,(v)$ are arbitrary functions.

It can be shown that for solutions of (3.4) with $N$ given by (3.6),
the equations (3.2$abc$) are automatically integrable.

Solutions of (3.2), (3.3) and  (3.4) describing the collisions of
line-waves have to satisfy certain boundary conditions. Two types of such
boundary conditions are discussed in the literature about colliding
gravitational waves in D=4. First one was introduced at the birth
of the subject [9, 10] and corresponds to spacetimes with flat initial
region I ($u<0$, $v<0$), two regions II and III ($0<u<u_0$, $v<0$ resp.
$0<v<v_0$, $u<0$) with metrics of the single line-wave solution and
an interacting region IV where the waves interact (Fig.3a).
The complete solution $K,N,\phi$ then has to fulfil prescribed
junction conditions on the wave-fronts $u$=0 and $v$=0 respectively.

The second type of boundary conditions was introduced recently by
Hayward [15] where the semiinfinite flat piece I of spacetime is excluded
and the waves are supposed to interact for all times with prescribed
asymptotic behaviour in the past timelike and null infinities (Fig.3b).
Such a prescription obviously avoids problems with junction conditions
on the wave-fronts and may be considered even physically more realistic.
We shall discuss solutions of the second type in D=3.
All functions $K,N,\phi$ are supposed to be smooth and to satisfy
following asymptotic conditions
$$(K,N,\phi)(u\to{-\infty},v\to{-\infty})=0\eqno(3.7a)$$
$$(K,N,\phi)(u\to{-\infty},v)=(K,N,\phi)(v)\eqno(3.7b)$$
$$(K,N,\phi)(u,v\to{-\infty})=(K,N,\phi)(u)\eqno(3.7c)$$

Rescaling $u$ and $v$ by arbitrary functions of $u$ and $v$ respectively,
$N$ and $\phi$ remain unchanged and $K$ changes additively. Hence coordinate
freedom is encoded in $K$ and (3.7$a$) fixes the coordinates.
Then (3.7$b$) and (3.7$c$) subject to constraints (3.2$ab$) play the role of
initial data.

Nontrivial solutions of equations (3.2), (3.3) and (3.4) can be found easily
by suitable dimensional reduction of vacuum solutions describing the
collisions of gravitational waves. Indeed, taking D=4 ansatz \footnote{We
consider the case of colliding waves with aligned polarisations,
i.e. without the term proportional to ${\rm d}x{\rm d}y$.} [15]
$${\rm d}s^2=-2{\rm e}^{-M}{\rm d}u\ {\rm d}v+{\rm e}^{-P}
({\rm e}^Q{\rm d}x^2+{\rm e}^{-Q}{\rm d}y^2)\eqno(3.8)$$
the D=4 Einstein equations read [15]
$$2P_{uu}-P^2_u+2M_uP_u-Q^2_u=0\eqno(3.9a)$$
$$2P_{vv}-P^2_v+2M_vP_v-Q^2_v=0\eqno(3.9b)$$
$$2M_{uv}+P_uP_v-Q_uQ_v=0\eqno(3.9c)$$
$$P_{uv}-P_uP_v=0\eqno(3.10)$$
$$2Q_{uv}-P_uQ_v-Q_uP_v=0\eqno(3.11)$$

It is easy to see that the equations (3.9), (3.10) and (3.11) give exactly
equations (3.2), (3.3) and (3.4) respectively after taking
$$P={N\over 2},\  M=K-{N\over 4},\  Q=\sqrt{2\kappa}\ \phi\eqno(3.12)$$

Moreover the D=4 asymptotic conditions [15]
$$(M,P,Q)(u\to{-\infty},v\to{-\infty})=0\eqno(3.13a)$$
$$(M,P,Q)(u\to{-\infty},v)=(M,P,Q)(v)\eqno(3.13b)$$
$$(M,P,Q)(u,v\to{-\infty})=(M,P,Q)(u)\eqno(3.13c)$$
give, after the substitution (3.12), the D=3 asymptotic conditions (3.7).

Hayward [15] has shown how to obtain easily solutions of the Einstein
equations (3.9), (3.10) and (3.11) fulfilling the asymptotic conditions (3.13)
knowing the solutions satisfying
``standard" boundary conditions (i.e. with semiinfinite flat wedge to the
past).
We feel that he made an important contribution carefully examining criteria
for regularity of incoming waves. Without this information one could not
exclude the possibility that the generic curvature singularities are caused
simply by singular incoming waves. He has shown, in fact, that the creation
of the curvature singularities to the future is generic even for the
regular incoming waves \footnote{The criteria of the ``incoming" regularity
in D=4 turned out to be rather restrictive, e.g. all solutions given by
Szekeres [11] have singular incoming waves, except the single Khan-Penrose
solution describing the collisions of the impulsive waves. The incoming
singularities for the Szekeres' family of solutions were studied by
Konkowski and Helliwell [17].}, though, in some special cases,
the formation of the singularity can be avoided.
In our paper we shall study the same set of solutions from the D=3 point
of view. In particular, we shall identify the criteria for the ``incoming"
regularity which turns out to be even more restrictive than D=4 criteria.
We evaluate also D=3 conditions for avoiding the formation of the final
singularities and show that for colliding regular incoming waves the
final singularity is inevitable.

\vskip 1 cm
\large
\S 4. Solutions of field equations.

\vskip 1 cm
\normalsize
The general solution for $\phi$ fulfilling (3.4) can be written in the form
[14, 15]
$$\phi=k\ln(1-f-g)+p\cosh^{-1}\Bigl[{1+f-g\over {1-f-g}}\Bigr]
+q\cosh^{-1}\Bigl[{1-f+g\over {1-f-g}}\Bigr]$$
$$+\int^\infty_0\bigl[A(\omega)J_0\bigl(\omega(1-f-g)\bigr)+B(\omega)N_0
\bigl(\omega(1-f-g)\bigr)\bigr]\sin \bigl(\omega(f-g)\bigr)\,{\rm d}\omega
\eqno(4.1)$$
where the functions $f(u)$ and $g\,(v)$ were introduced in (3.6),
$k,p,q$ are real numbers, the amplitudes $A(\omega),\ B(\omega)$ may
be integrable functions or distributions and $J_0$ and $N_0$ are the
zero-order Bessel and Neumann function respectively.

The different terms in (4.1) possess different behaviour when
$f+g\to {1^-}$. The first term is obviously singular, so are the second and
third terms which are usually referred to as gravitational
solitons. The fourth term is regular but the fifth one again contributes
to the divergence due to the behaviour of the Neumann functions near zero.
We have to elucidate what restrictions on the general solution (4.1) come
from the asymptotic conditions (3.7).

We choose, obviously, the functions $f(u)$ and $g\,(v)$ such that
$$f(u=-\infty)=g\,(v=-\infty)=0.\eqno(4.2)$$
Consider now Eq. (3.2$a$) at $v$=$-\infty$, i.e.
$${f_{uu}\over {1-f}}+{f_u\over {1-f}}K_u=\kappa\phi_u\phi_u\eqno(4.3)$$
hence
$${f_{uu}\over {f^2_u}}=\kappa\phi_f\phi_f(1-f)-K_f.\eqno(4.4)$$

We assume that the function $f(u)\ \bigl(g\,(v)\bigr)$ is strictly monotonic.
The asymptotic behaviour of $\phi_f$ for $f\sim 0$ (and $g$=0) reads [15]
$$\phi_f\sim{p\over {\sqrt f}}-C+p\sqrt f+...\eqno(4.5)$$
where
$$C=k-\int^\infty_0\omega\bigl[A(\omega)J_0(\omega)+B(\omega)N_0(\omega)\bigr]
{\rm d}\omega.\eqno(4.6)$$

Setting
$${f_{uu}\over {f^2_u}}=H(f)\eqno(4.7)$$
where $H$ is an appropriate function, we may integrate Eq. (4.4)
to obtain behaviour of $K$ for $f\sim 0$, i.e.
integrate
$$K_f=-H(f)+\kappa\phi_f\phi_f(1-f).\eqno(4.8)$$

We have for $f\sim 0$
$$\kappa\phi_f\phi_f(1-f)\sim \kappa \Bigl[{p^2\over f}+C^2+p^2
-{2pC\over {\sqrt f}}+...\Bigr]\eqno(4.9)$$

To ensure regular behaviour of $K$ (or to compensate the term
$\kappa p^2\over f$) we must have
$$H(f)=\widetilde H(f)+{\kappa p^2\over f}\eqno(4.10)$$
where
$$\lim_{f\to 0}f\widetilde H(f)=0.\eqno(4.11)$$

Starting with $\widetilde H(f)$=0, we have to solve the equation
$${f_{uu}\over {f^2_u}}={\kappa p^2\over f}\eqno(4.12)$$
with the boundary condition $f(-\infty)$=0.

The latter condition excludes the cases $\kappa p^2<1$; hence, we have
the following class of  solutions\footnote{We may also understand
the equations (4.13) and (4.14) as a pure choice of coordinates,
equivalent to fixing $K$ on $g=0$ and $f=0$.}
$$f(u)=[-a\,(u-u_S)]^{1/(1-\kappa p^2)},\
{\rm for}\ \kappa p^2>1\eqno(4.13a)$$
$$f(u)=\exp[a\,(u-u_S)],\ {\rm for}\ \kappa p^2=1\eqno (4.13b)$$
where $u_S$ is arbitrary and $a$ is a positive number.

In an analogous way we obtain\footnote{The choices (4.13b) and (4.14b)
slightly enlarge the possible coordinate fixing
of D=4 solutions obtained by Hayward [15].}
$$g\,(v)=[-b\,(v-v_S)]^{1/(1-\kappa q^2)},\
{\rm for}\ \kappa q^2>1\eqno(4.14a)$$
$$g\,(v)=\exp[b\,(v-v_S)],\ {\rm for}\ \kappa q^2=1.\eqno (4.14b)$$

We can conclude that for the choices of $f(u)$ and $g\,(v)$ (4.13) and (4.14)
respectively, the presence of the solitonic terms in the general solution
(4.1) is inevitable in order to fulfil the asymptotic conditions (3.7).

As is well-known, the solitonic terms are necessary in D=4 for standard
colliding waves metrics in order to satisfy the junction conditions
on the wave-fronts $u$=0 and $v$=0 [14]. In our case, however, we do not
need to cope with the junction conditions. One may therefore wonder
whether the solitonic terms are really necessary in order to satisfy
only the asymptotic conditions (3.7). Though we did not attempt to solve
the problem in full generality, being satisfied with large classes
(4.13) and (4.14) of allowed $f(u)$ a $g\,(v)$, we give, nevertheless,
an indication that the solitonic terms $are$ needed. Let us put $p^2$=0
in (4.9) and (4.10). Thus
$$K_f=-\widetilde H(f)+\kappa C^2+...\eqno(4.15)$$
where $\widetilde H(f)$ satisfies condition (4.11). We choose for concreteness
a power-like behaviour $(\alpha>0)$
$$\widetilde H(f)=f^{-1+\alpha}\eqno(4.16)$$
and show that the condition $f(-\infty)$=0 is necessarily violated.
We write
$${f_{uu}\over {f^2_u}}=f^{-1+\alpha}\eqno(4.17)$$
hence
$$f_u=D\ \exp\Bigl({1\over \alpha}f^\alpha\Bigr).\eqno(4.18)$$

Making a substitution
$$r={1\over \alpha}f^\alpha\eqno(4.19)$$
we arrive at
$$r_u=D\,(\alpha r)^{1-{1\over \alpha}}\,{\rm e}^r\eqno(4.20)$$
thus
$$D\,\alpha^{1-{1\over\alpha}}\int {\rm d}u=\int {\rm d}r\
r^{{1\over \alpha}-1}\,{\rm e}^{-r}.\eqno(4.21)$$

If $f(-\infty)$=0, we have
$$D\,\alpha^{1-{1\over\alpha}}\int ^u_{-\infty}{\rm d}u'=\int^r_0{\rm d}r'\
(r')^{{1\over \alpha}-1}\,{\rm e}^{-r'}.\eqno(4.22)$$

Clearly the incomplete $\Gamma$-function is finite while the
left-hand-side is infinite. Thus $f(-\infty)$ cannot be equal to zero.

We complete our discussion by identifying the remaining unknown function
$K(u,v)$. Taking the general solution (4.1), the integration of constraints
(3.2) cannot in general be performed explicitly, nevertheless, the processes
of singularity formation $can$ be studied (see \S 6). In particular case,
however, when the Fourier-Bessel modes are absent, we may give $K(u,v)$
even explicitly. It reads
$$K=-\kappa \bigl(k^2-(p+q)^2\bigr)\ln \bigl(1-f-g\bigr)
+\kappa q^2\ln \bigl(1-f\bigr)$$
$$+\kappa p^2\ln \bigl(1-g\bigr)+4\kappa pq\ln
\bigl(\sqrt {fg}+\sqrt {1-f}\sqrt {1-g}\bigr)\eqno(4.23)$$
$$-2\kappa kp\ln \biggl({{\sqrt {1-g}+\sqrt {f}}\over
{\sqrt {1-g}-\sqrt {f}}}\biggr)-2\kappa kq\ln \biggl(
{{\sqrt{1-f}+\sqrt {g}}\over {\sqrt {1-f}-\sqrt {g}}}\biggr)$$

where $f(u)$ and $g\,(v)$ are given by (4.13) and (4.14) respectively.
By construction this expression satisfies the asymptotic condictions (3.7).

\vskip 1 cm
\large
\S 5. Criteria for ``incoming" regularity.

\vskip 1 cm
\normalsize
We have identified in the previous sections a large class of solutions
of the Einstein equations (3.2-4) in D=3 satisfying the asymptotic conditions
(3.7). As we have already mentioned the asymptotic colliding line-waves
interacting for all times seem better to correspond to physical reality than
colliding waves with sharp wave-fronts. A question remains, however,
whether all such solutions (4.1) with $f$ and $g$ given by (4.13) and
(4.14) are ``enough" physically realistic. Though D=3 is itself not very
physical dimension it makes sence to attempt a formulation of some
criteria ensuring an ``acceptability" of the process from D=3 point of
view.

In the D=4 case Hayward [15] has postulated a criterion that parallelly
propagated curvature singularities in the asymptotic caustics
$f$=1, $g$=0 and $f$=0, $g$=1 should be absent. He has also shown that such
a criterion amounts to boundedness of the amplitudes of the incoming
gravitational waves in Brinkmann coordinates in the asymptotic caustics.
We shall also claim in D=3 that parallelly propagated curvature singularities
should be absent and show that this again amounts to the boundedness of the
metric in Brinkmann coordinates. We note, nevertheless, that points in
past null infinities marked by jagged lines in Fig.3b are in this case
quasiregular but still singular points of the manifold.

We first identify the criterion for boundedness of the metric components
in Brinkmann coordinates and pick up, for concreteness, the case $g$=0.
In this null infinity the metric looks as follows
$${\rm d}s^2=-{\rm e}^{-K(u,v=-\infty )}{\rm d}u{\rm d}v+
{\rm e}^{-N(u,v=-\infty )}{\rm d}x^2.\eqno(5.1)$$
We make the transformation of the coordinate $u$ setting
$${{\rm d}u'\over {{\rm d}u}}={\rm e}^{-K(u)}.\eqno(5.2)$$
Thus
$${\rm d}s^2=-{\rm d}u'{\rm d}v+{\rm e}^{-N(u')}{\rm d}x^2
\eqno(5.3)$$
and, following (2.5),
$$h(u')={\rm e}^{{N(u')}\over 2}{{{\rm d}^2}\over {{\rm d}
u'^2}}{\rm e}^{-{{N(u')}\over 2}}.\eqno(5.4)$$

We rewrite (5.4) in the original coordinates
$$h(u)={{{\rm e}^{2K}}\over 2}\bigl({{N^2_u}\over 2}-N_{uu}-N_uK_u\bigr)
=-{\rm e}^{2K}\kappa \phi_u\phi_u\eqno(5.5)$$
where the last equality follows from the Einstein equation (3.2$a$).

We have to study behaviour of $h(u)$ at $f(u)$=1. We first recall the
asymptotic behaviour of the Bessel and Neumann functions for $w\to 0^+$
$$J_0(w)\sim 1-{w^2\over 4}+...\ \ \ \ \ \ \ $$
$$J'_0(w)\sim -{w\over 2}+...\ \ \ \ \ \ \ \ \ \ \ $$
$$N_0(w)\sim (1-{w^2\over 4})\ln w+...\eqno(5.6)$$
$$N'_0(w)\sim {1\over w}-{w\over 2}\ln w+...\ \ $$

The scalar field solution (4.1) gives for $g$=0
$$\phi =k\ln(1-f)+p\ln{1+\sqrt f\over 1-\sqrt f}$$
$$+\int^\infty_0\bigl[A(\omega)J_0\bigl(\omega (1-f)\bigr)
+B(\omega)N_0\bigl(\omega(1-f)\bigr)\bigr]\sin(\omega f)\,
{\rm d}\omega\eqno(5.7)$$

Then it follows from (5.6) and (5.7) the behaviour of $\phi_u$ for $f\sim 1$
$$\phi_u\sim {cf_u\over 1-f}+d\, f_u\ln (1-f)+ef_u+hf_u(1-f)\ln (1-f)+...
\eqno(5.8)$$
where
\begin{eqnarray}
c&=&p-k-\int^\infty_0B(\omega)\sin (\omega)\,{\rm d}\omega\nonumber\\
d&=&\int^\infty_0\omega B(\omega)\cos (\omega)\,{\rm d}\omega\nonumber\\
e&=&{p\over 2}+\int^\infty_0\omega\bigl[A(\omega)+B(\omega)\ln\omega\bigr]
\cos(\omega)\,{\rm d}\omega\nonumber\\
h&=&{1\over 2}\int^\infty_0\omega^2B(\omega)\sin(\omega)\,{\rm d}\omega
\nonumber\\\nonumber
\end{eqnarray}
Integrating the constraint (3.2$a$) for $f\sim 1$ we have (for the choice
(4.13)\,)
$$K_u={\kappa c^2f_u\over 1-f}+2\kappa cd\,f_u\ln (1-f)+...\eqno(5.9)$$
hence
$$K=-\kappa c^2\ln (1-f)+{\rm bounded}\eqno(5.10)$$

If $c\ne 0$ then $h(u)$ given by (5.5) is obviously singular. We have,
therefore, the first necessary condition of boundedness of the metric
in Brinkmann coordinates. It reads
$$c=p-k-\int^\infty_0B(\omega)\sin (\omega)\,{\rm d}\omega=0.$$

With $c$=0 we have
$$h(u)=-\kappa \ {\rm const}\ f^2_u\bigl[d^2\ln^2(1-f)+2de\ln(1-f)+
{\rm bounded}\bigr].\eqno(5.11)$$

Thus conditions for boundedness of $h(u)$ read
$$c=d=0.\eqno(5.12a)$$

In the asymptotic caustic $f$=0, $g$=1, we require analogously
$$c'=d=0\eqno(5.12b)$$
where
$$c'=q-k+\int^\infty_0B(\omega)\sin(\omega)\,{\rm d}\omega.$$

We turn now to criteria of absence of parallelly propagated singularities
in the asymptotic caustics $f$=0, $g$=1 and $f$=1, $g$=0.
We shall work with the case $g$=0 and we choose the coordinates given by
(5.2). The following is the parallelly propagated orthonormal frame for
geodesics respecting $x$-symmetry
$$m=(1,1,0),\ n=(1,-1,0),\ l=(0,0,{\rm e}^{N(u')\over 2}).\eqno(5.13)$$
The only nonzero components of the Riemann tensor in this frame are given by
$$R_{lmlm}=R_{lmln}=R_{lnln}=\kappa \phi_{u'}\phi_{u'}\eqno(5.14)$$

Turning back to the original coordinates we have the condition for absence of
the parallelly propagated singularity as follows:
$$\kappa {\rm e}^{2K}\phi_u\phi_u={\rm bounded.}\eqno(5.15)$$

We observe that this is the already encountered condition of boundedness
of the metric in Brinkmann coordinates, therefore, the conditions
(5.12) are also conditions for absence of the parallelly
propagated singularities in the asymptotic caustics.
We note that the conditions $c$=$c'$=0
imply boundedness of the scalar field itself as follows from (5.7)
and from the asymptotic behaviour  of the Bessel and Neumann functions
given by (5.6). Since $\phi$ is the scalar field, its boundedness
should have been expected.

We note also that in our particular case we may give a sufficient condition
for boundedness of all covariant derivatives of the Riemann tensor in the
frame (5.13). Indeed, we use the formula valid in D=3 [1]
$$R_{\alpha\beta\gamma\delta}=\kappa\bigl[(g_{\alpha\gamma}T_{\beta\delta}
+g_{\beta\delta}T_{\alpha\gamma}-g_{\alpha\delta}T_{\beta\gamma}-
g_{\beta\gamma}T_{\alpha\delta})$$
$$+T(g_{\alpha\delta}g_{\beta\gamma}-g_{\alpha\gamma}g_{\beta\delta})\bigr]
\eqno(5.16)$$
where $T_{\alpha\beta}$ is the matter stress tensor.

Taking into account that the only nonzero component of the $T_{\alpha\beta}$
in our case is
$$T_{u'u'}=\phi_{u'}\phi_{u'}\eqno(5.17)$$
and the only nonzero Christoffel symbols are
$$\Gamma^x_{xu'}=-{1\over 2}N_{u'},\ \Gamma^v_{xx}=-N_{u'}\ {\rm e}^{-N}
\eqno(5.18)$$
we arrive after some work at the sufficient condition of boundedness of all
covariant derivatives of the Riemann tensor in the frame (5.13) up to the
order $J$.

It reads
$${{\rm d}^j\phi(u')\over {\rm d}u'^j}={\rm bounded,\ for\ all\ }j\le J+1.
\eqno(5.19)$$

Written in the original coordinates
$$\Bigl({\rm e}^K{{\rm d}\over {\rm d}u}\Bigr)^j\phi(u)={\rm bounded,\
for\ all\ }j\le J+1.\eqno(5.20)$$

If $J$=0 we recover
$${\rm e}^K\phi_u={\rm bounded}\eqno(5.21)$$
which clearly implies (5.15).

We shall not present the general discussion of the conditions (5.19) for
the solutions (4.1). We note, however, that the background solution
with $B(\omega)$=$A(\omega)$=0 and \it p=q=k \rm gives the bounded all orders
covariant derivatives of the Riemann tensor evaluated in the parallelly
propagated orthonormal frame (5.13).

Conditions (5.12) are obviously satisfied by huge number
of solutions (4.1) (e.g. if $B$=0 and $p=k=q,\  A(\omega)$ can be
arbitrary). In the next paragraph we shall study which of them lead to
the formation of the final curvature singularities.

\vskip 1 cm
\large
\S 6. Scalar curvature singularities.

\vskip 1 cm
\normalsize
Scalar curvature $R$ for the metric (3.1) reads
$$R=-4{\rm e}^KK_{uv}\eqno(6.1)$$
or
$$R=-4\kappa {\rm e}^Kf_ug_v\phi_f\phi_g\eqno(6.2)$$
where (6.2) follows from (6.1) and Einstein equation (3.2c). We wish to
investigate scalar curvature $R$ near the caustic
$$1-f(u)-g\,(v)=0.\eqno(6.3)$$
For this purpose we introduce coordinates
$$t=1-f(u)-g\,(v)\eqno(6.4)$$
$$z=f(u)-g\,(v)\eqno(6.5)$$
thus
$$R\sim {\rm e}^K(\phi^2_t-\phi^2_z)\eqno(6.6)$$
since $f_u$ and $g_v$ are bounded at $t$=0.

{}From the form of the general solution (4.1) and asymptotic behaviour
of the Bessel and Neumann functions (5.6) it follows for $t\sim 0$
(see also [15])
$$\phi_t\sim {E\over t}-{Ft\over 2}+...\eqno(6.7)$$
$$\phi_z\sim G\ln t+I+...\eqno(6.8)$$
where
$$E=k-p-q+\int^\infty_0B(\omega)\sin(\omega z)\,{\rm d}\omega\eqno(6.9a)$$
$$F=p\ (1+z)^{-2}+q\ (1-z)^{-2}+{1\over 2}\int^\infty_0
\omega^2[2A(\omega)-B(\omega)]\sin(\omega z)\,{\rm d}\omega\eqno(6.9b)$$
$$G=\int^\infty_0\omega B(\omega)\cos(\omega z)\,{\rm d}\omega\eqno(6.9c)$$
$$I=p\ (1+z)^{-1}-q\ (1-z)^{-1}+\int^\infty_0\omega[A(\omega)+B(\omega)
\ln\omega]\cos(\omega z)\,{\rm d}\omega.\eqno(6.9d)$$

The asymptotic behaviour of $K$ for $t\sim 0$ can be obtained by integrating
the Einstein equations (3.2)
$$K_f=\kappa t\phi^2_f-{f_{uu}\over f^2_u}\eqno(6.10a)$$
$$K_g=\kappa t\phi^2_g-{g_{vv}\over g^2_v}\eqno(6.10b)$$
hence
$$K_t=-\kappa t(\phi^2_t+\phi^2_z)+{1\over 2}\Bigl({f_{uu}\over f^2_u}+
{g_{vv}\over g^2_v}\Bigr)\eqno(6.11a)$$
$$K_z=-2\kappa t\phi_t\phi_z+{1\over 2}\Bigl({g_{vv}\over g^2_v}-
{f_{uu}\over f^2_u}\Bigr)\eqno(6.11b)$$

The second terms in (6.11) for the choice (4.13) have no influence
on divergence of the scalar curvature. Using (6.7) and (6.8) we arrive at
$$K=-\kappa E^2\ln t+...\eqno(6.12)$$

Collecting (6.7), (6.8) and (6.12) we have
$$R\sim t^{-\kappa E^2}\ \Bigl[{E^2\over t^2}-G^2\ln^2t+...\Bigr]\eqno(6.13)$$

We conclude that $R$ can be nonsingular on the caustic $t$=0 only if \it
E=G=\rm 0. But both $E$ and $G$ have nontrivial dependence on $z$ unless
$B(\omega)$=0. In this case $G$=0 automatically and $E$=0 means
$k-p-q=0$. Thus conditions for avoiding the scalar curvature
singularities on the caustic $t$=0 ($z\ne\pm 1$) read
$$B(\omega)=0,\ k-p-q=0.\eqno(6.14)$$

Comparing with conditions of ``incoming" regularity (5.12) we see
that for regular incoming waves the scalar curvature singularities at the
caustic $t$=0 are inevitable.

\vskip 1 cm
\large
\S 7. Conclusions and outlook.

\vskip 1 cm
\normalsize
We have obtained the large class of solutions of equations of motion for
self-gravitating scalar field in D=3. They describe scattering
of excitations and corresponding spacetimes are inevitably singular
to the future for all incoming waves free of parallelly propagated
singularities.
The incoming waves do not have sharp wave-fronts, they interact
for all times, the spacetime being flat to the past only asymptotically.
In distinction to the case of D=4 vacua, by dimensional reduction of which
our family of spacetimes was constructed, there are no exceptions from
the final singularity formation.

The class of considered solutions possesses an interesting property.
The solutions are given as the Fourier-Bessel mode decompositions around
the background soliton. The asymptotic flatness gives restriction
on the amplitudes $p$, $q$ of the solitary waves. Namely
$$p^2\kappa\ge 1,\ q^2\kappa\ge 1\eqno(7.1)$$

\noindent It means that we cannot switch off gravitational interaction by
setting $\kappa$=0 without violating the conditions (7.1) and, consequently,
the proper asymptotic behaviour. We see also
that the phase space of scalar field in D=3 is enlarged
due to proper self-gravitation.

The existence of Fourier-Bessel decomposition of general solution is more
interesting in D=3 than in D=4. Indeed, an existence of a mode decomposition
of scalar field solutions is more appealing than a mode decomposition
of a metric component. As we have already mentioned in the Introduction,
the matter field degrees of freedom are the only degrees of freedom in
D=3 and we succeeded to extract them rather
explicitly. The next logical step would rely on quatizing the matter degrees
of freedom. We did not attempt to do it in this paper but we provide
yet two more comments on the subject. First of all, we restricted ourselves
to solutions with a line symmetry. It may well happen that the singularity
formations come from high symmetry of the problem and they would
disappear when  dealing with waves with  a finite transverse size [8, 18, 19].
This problem, to our knowledge, remains open also in D=4 and in D=3 it would
probably influence the proper quantum treatment. Second comment relies
on the fact that the background solution, around which the mode
decomposition is made, is itself singular; hence, unless something
surprising happens, one should not probably expect the smearing of
singularities in the quantum picture. Nevertheless, the quantum information
about the role and behaviour of the scattering singularities is certainly
of an interest and we hope to return to the problem elsewhere.

\vskip 1 cm
\large

Acknowledgment.

\vskip 1 cm
\normalsize
We are grateful to Ji\v r\' \i \  Bi\v c\' ak for comments on the manuscript.
We wish to thank also H Bou\v skov\' a and G Han\' akov\' a
for drawing the figures.

\newpage
\large

References.

\normalsize
\vskip 1 cm
\noindent \, [1] Giddings S, Abbot J and Kucha\v r K 1984 \it Gen. Rel. Grav.
\rm {\bf 16} 751

\noindent \, [2] Deser S, Jackiw R and 't Hooft G 1984 \it Ann. Phys. (N.Y.)
\rm {\bf 152} 220

\noindent \, [3] 't Hooft G 1988 \it Commun. Math. Phys. \rm {\bf 117} 685

\noindent \, [4] Deser S and Jackiw R 1988 \it Commun. Math. Phys. \rm
{\bf 118} 495

\noindent \, [5] Witten E 1989 \it Nucl. Phys. \rm B {\bf 323} 113

\noindent \, [6] Carlip S 1989 \it Nucl. Phys. \rm B {\bf 324} 106

\noindent \, [7] Lancaster D Kyoto preprint KUNS 1016, HE(TH) 90/06 (May
1990)

\noindent \, [8] Garriga J and Verdaguer E 1991 \it Phys. Rev. \rm D
{\bf 43} 391

\noindent \, [9] Szekeres P 1970 \it Nature \rm {\bf 228} 1183

\noindent [10] Khan K A and Penrose R 1971 \it Nature \rm {\bf 229}
185

\noindent [11] Szekeres P 1972 \it J. Math. Phys. \rm {\bf 13} 286

\noindent [12] Matzner R A and Tipler F J 1984 \it Phys. Rev. \rm D
{\bf 29} 1575

\noindent [13] Ferrari V and Iba\~ nez J 1989 \it Class. Quantum Grav. \rm
{\bf 6} 1805

\noindent [14] Feinstein A and Iba\~ nez J 1989 \it Phys. Rev. \rm D
{\bf 39} 470

\noindent [15] Hayward S A 1990 \it Class. Quantum Grav. \rm {\bf 7} 1117

\noindent [16] Penrose R 1965 \it Rev. Mod. Phys. \rm {\bf 37} 215

\noindent [17] Konkowski D A and Helliwell T M 1989 \it Class. Quantum
Grav. \rm {\bf 6} 1847

\noindent [18] Klim\v c\'\i k C 1988 \it Phys. Lett. \rm {\bf 208B} 373

\noindent [19] Yurtsever U 1988 \it Phys. Rev. \rm D {\bf 38} 1731
\newpage

\large
Figure Captions.
\vskip 1 cm
\normalsize

Fig.1a: Wave-front $v_w$(=0) in the $u$-wave Rosen coordinates with generating
geodesics.

\vskip 1 cm
Fig.1b: Wave-front $v_w$(=0) in the $u$-wave Brinkmann coordinates.
The wave-front of the $u$-wave is also given by $u_w$=0.

\vskip 1 cm
Fig.2: The $u$-wave in Rosen coordinates with a part, indicated by hatch,
cut out by the crossing wave-front. The jagged line indicates the coordinate
singularity $u$=$u_0$. The $x$-coordinate is suppressed.

\vskip 1 cm
Fig.3a: Expected shape of the spacetime describing the collision of two
waves with sharp wave-fronts $u_w$=0 and $v_w$=0. The jagged lines indicate
the singular points of spacetime.

\vskip 1 cm
Fig.3b: Expected shape of the spacetime describing the collision of two
waves without sharp wave-fronts. The jagged lines indicate (asymptotically)
the singular points of the spacetime.

\end{document}